\documentclass[prc,aps,preprint]{revtex4-1}
\usepackage{graphicx}
\usepackage{natbib}
\usepackage{subfig}
\usepackage{float}
\begin{document}

\title{Quasi-dynamical symmetries in the backbending of  chromium isotopes}

\author{Ra\'ul A. Herrera}
\affiliation{Department of Physics and \\
Center for Astrophysics and Space Sciences, 
University of California San Diego,
9500 Gilman Drive, La Jolla, CA 92093} 
\author{Calvin W. Johnson}
\affiliation{Department of Physics, San Diego State University,
5500 Campanile Drive, San Diego, CA 92182}
\affiliation{Center for Astrophysics and Space Sciences, 
University of California San Diego,
    9500 Gilman Drive, La Jolla, CA 92093 }


\begin{abstract}
\begin{description}
\item[Background]  
Symmetries are a powerful way to characterize nuclear wave functions.  
A true dynamical symmetry,  where the Hamiltonian is block-diagonal in subspaces defined by the group, is rare.  More likely is a \textit{quasi-dynamical} symmetry:  states with different quantum numbers 
 (i.e. angular momentum) nonetheless sharing similar group-theoretical decompositions. 
\item[Purpose]    
We use group-theoretical decomposition to investigate backbending, an abrupt change in the moment of inertia along the yrast line, in
 $^{48,49,50}$Cr:
prior mean-field calculations of these nuclides suggest a change from strongly prolate  
to more spherical configurations as one crosses the backbending and increases in angular momentum.  
\item[Methods]    
We  decompose configuration-interaction shell-model wavefunctions 
using 
  the  SU(2) groups $L$ (total orbital angular momentum) and 
$S$ (total spin), and  the groups SU(3) and SU(4).  We do not need  a special basis but  only matrix elements of  Casimir operators, applied with a modified Lanczos algorithm.
 \item[Results]     
We find  quasi-dynamical symmetries, albeit often of a different 
character above and below the backbending, for each group. 
While  the strongest evolution was  in SU(3),  the decompositions did not suggest a decrease in deformation. We point out with a simple example
that mean-field and SU(3) configurations may give very different pictures of deformation. 
\item[Conclusions] 
Persistent quasi-dynamical symmetries for several groups allow us to identify the members of a band and to characterize 
how they evolve with increasing angular momentum, especially before and after backbending. 
\end{description}
\end{abstract}

\maketitle

\section{Introduction}

\begin{figure}
\centering
\includegraphics[scale=0.55,clip]{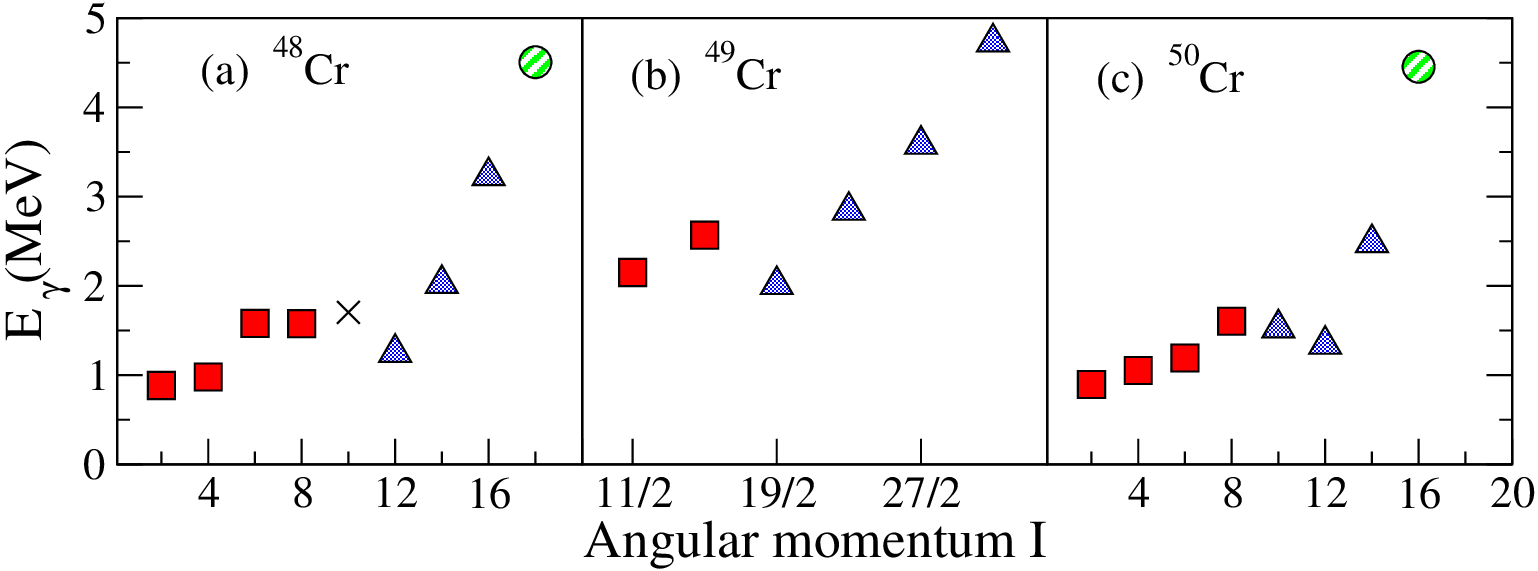}
\caption{(Color online) 
Backbending in $^{48-50}$Cr, as signaled by the evolution of $E_\gamma(I) = E(I)-E(I-2)$.  The distinct shapes/colors represent, to the best of our ability to identify, 
different configurations along the yrast as discussed in detail in the text: (red) solid squares for the lower sub-band, (blue) dotted triangles for the upper sub-band, 
and a black `x' and (green) striped circle for upper and lower `intruder' levels, respectively.   The calculated values are in good agreement with experiment (not shown).}
\label{BackBending}
\end{figure}

Backbending is an abrupt change in the nuclear moment of inertia along the yrast line 
\cite{ring2004nuclear}, seen in nuclides ranging from $^{22}$Ne \cite{PhysRevLett.42.622} through the 
actinides \cite{PhysRevLett.51.1522}. 
In a rotational band with constant moment of inertia the gamma transition energy $E_\gamma(I) = E(I) - E(I-2)$ grows steadily with 
angular momentum $I$, but in backbending $E_\gamma(I)$ abruptly falls and then rises again with a different slope,
as illustrated in Fig.~\ref{BackBending} for $^{48,49,50}$Cr.

There are three general explanations for the change in the moment of inertia \cite{ring2004nuclear}

\noindent $\bullet$ a change in deformation;

\noindent $\bullet$ a change from superfluid to normal phase;

\noindent $\bullet$ a change in alignment of quasiparticles.

Of course, backbending may be due to a mixture of these explanations; furthermore, it may not be the same for 
all nuclei \cite{PhysRevC.58.2765}.

Because backbending occurs mostly frequently in heavy nuclei, 
most calculations of backbending have used mean-field and related methods \cite{bengtsson1978conditions}, such as cranked  
Hartree-Fock-Bogoliubov \cite{sorensen1976treatment,cwiok1978discussion,sugawara1988g,PhysRevC.21.448} and the (angular-momentum) projected shell model \cite{velazquez1999backbending}.
A favorite target of theory, however, has been backbending in the chromium 
isotopes \cite{PhysRevC.49.1347,Cameron1996266,PhysRevLett.79.4349,PhysRevC.58.808,PhysRevC.56.1313,PhysRevC.58.808,PhysRevC.66.021302}, 
because in addition to mean-field and similar studies \cite{PhysRevC.58.2765,PhysRevLett.83.1922,velazquez2001band} one can fully diagonalize the nuclear Hamiltonian in the $1p$-$0f$ (`$pf$') shell using configuration-interaction methods
\cite{PhysRevLett.75.2466,PhysRevC.50.225,PhysRevC.53.188,PhysRevC.54.R2150,PhysRevC.55.187,PhysRevC.83.057303,PhysRevC.71.054316,PhysRevC.73.044327}.

We will discuss some of these prior investigations  in more detail below. We are especially motivated, however by recent assertions  \cite{PhysRevC.83.057303} 
that that for $^{48}$Cr 
 the lower sub-band (below the backbending) can be associated with a well-defined intrinsic state, 
but not the upper sub-band  (above the backbending).  We follow this up by decomposing the nuclear wavefunctions into subspaces defined by group Casimir operators,
that is, operators which are invariant under all elements of a Lie group and its related algebra \cite{chen1989group,talmi1993simple,rowe2010fundamentals}. 
We see strong characteristics of \textit{quasi-dynamical symmetry}, that is, consistent fragmentation of the wavefunction with increasing $I$; 
in most cases we see a change as one crosses the backbending, and in SU(3) we see 
significant evolution of the fragmentation in the upper sub-band as $I$ increases.

As described below in section \ref{decomposition}, we use an efficient method to decompose a wavefunction according to subspaces labeled by 
eigenvalues of Casimir operators. 
We choose total orbital angular momentum $L$ and total spin $S$, both of which belong to group the group SU(2),
as well as the groups SU(3), and SU(4). We limit ourselves to two-body Casimirs.

\section{MIcroscopic methods}

\subsection{Configuration-interaction shell model}

\begin{figure}
\centering
\includegraphics[scale=0.60,clip]{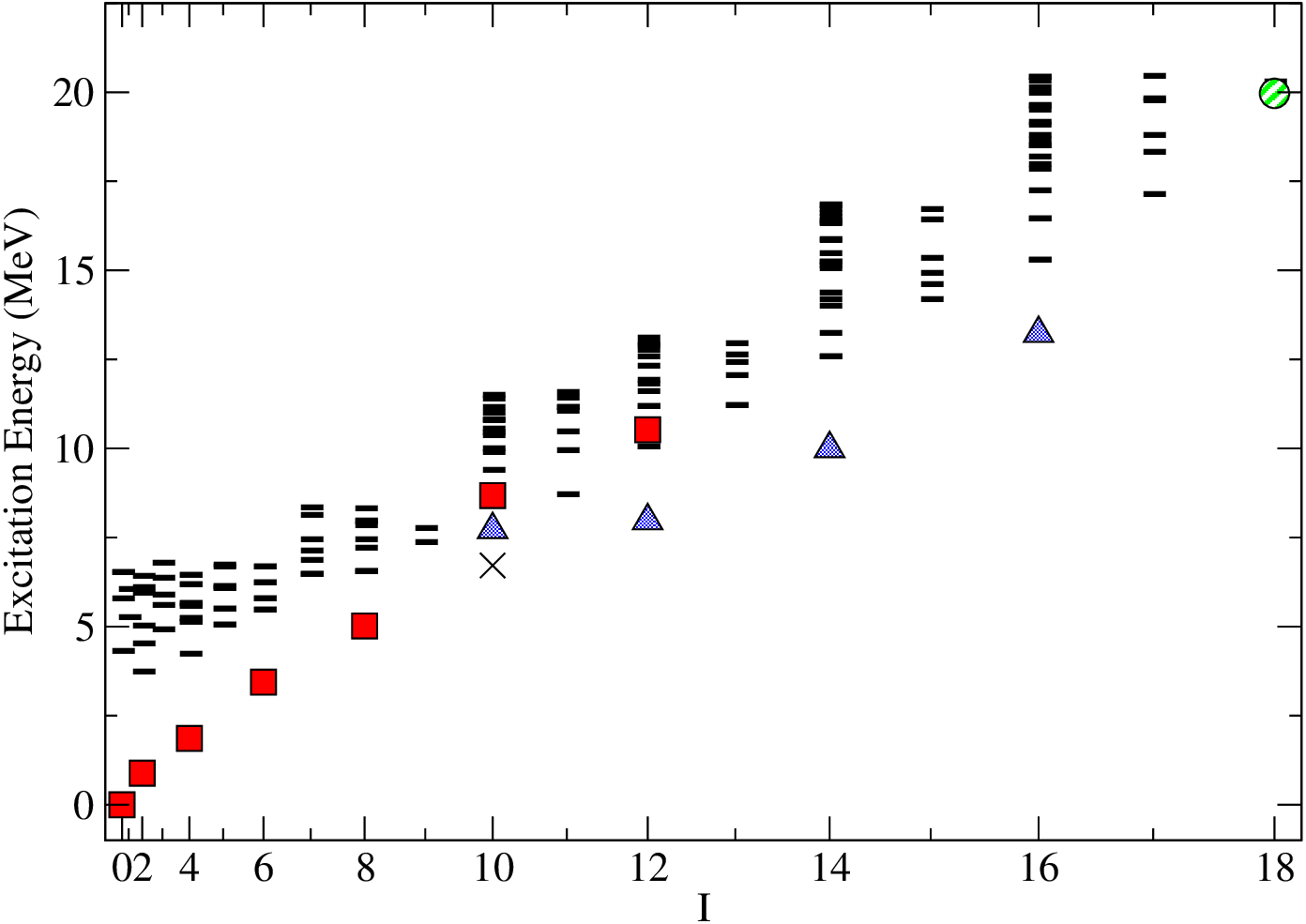}
\caption{(Color online) 
Calculated spectrum of $^{48}$Cr.  The $x$-axis (angular momentum $I$) is scaled as $I(I+1)$ so as to emphasize rotational bands.  The labeling of levels, i.e., 
(red) squares, (blue) triangles, and (green) circles, correspond to the same (initial) state as in Panel (a) of Fig.~\ref{BackBending}.  According to our decompositions,
the yrast state at $I=10$, marked by as `x,' belongs to neither the lower nor upper sub-bands.
 Bars indicate levels found in our 
calculation but which we do not decompose.}
\label{cr48spect}
\end{figure}

We carry out calculations in the framework of the configuration-interaction (CI) shell model \cite{BG77,br88,ca05}, which expresses the nuclear Hamiltonian as a large-dimensioned matrix 
in a basis of shell-model Slater determinants (antisymmetrized products of single-particle states), recasting the many-body Schr\"odinger equation as a matrix 
eigenvalue problem,
\begin{equation}
\hat{H} | \Psi_i \rangle = E_i | \Psi_i \rangle. \label{nucl_ham}
\end{equation}
We find the low-lying eigenpairs, 
via the Lanczos algorithm, using the {\tt BIGSTICK} configuration-interaction code \cite{BIGSTICK}. Because the Hamiltonian is rotationally 
invariant, the total magnetic quantum number $M$ (or $J_z$, the $z$ component of the total angular momentum) is conserved and one can easily construct a 
basis with fixed $M$; this is called an $M$-scheme basis. 

Although \textit{ab initio} calculations for $0p$-shell nuclides are now routine, for the chromium isotopes we use 
the modified G-matrix interaction for the $1p$-$0f$ ($pf$) shell  GXPF1  \cite{PhysRevC.65.061301}, which assumes a frozen $^{40}$Ca core and valence particles restricted to the $1p$-$0f$ single-particle space.  Like other high-quality semi-phenomenological interactions in the $pf$ shell, calculated spectra using GXPF1 have good agreement with 
experiment (which we do not show to avoid further cluttering our figures).  
We also made decompositions in the same space using the monopole-modified Kuo-Brown effective interaction version KB3G \cite{kb3g} and the modified GXPF1 interaction, version A, \cite{honma2005shell} 
and found very similar results.

\subsection{Group decomposition  and quasi-dynamical symmetry}

\label{decomposition}

\begin{figure}
\centering
\includegraphics[scale=0.60,clip]{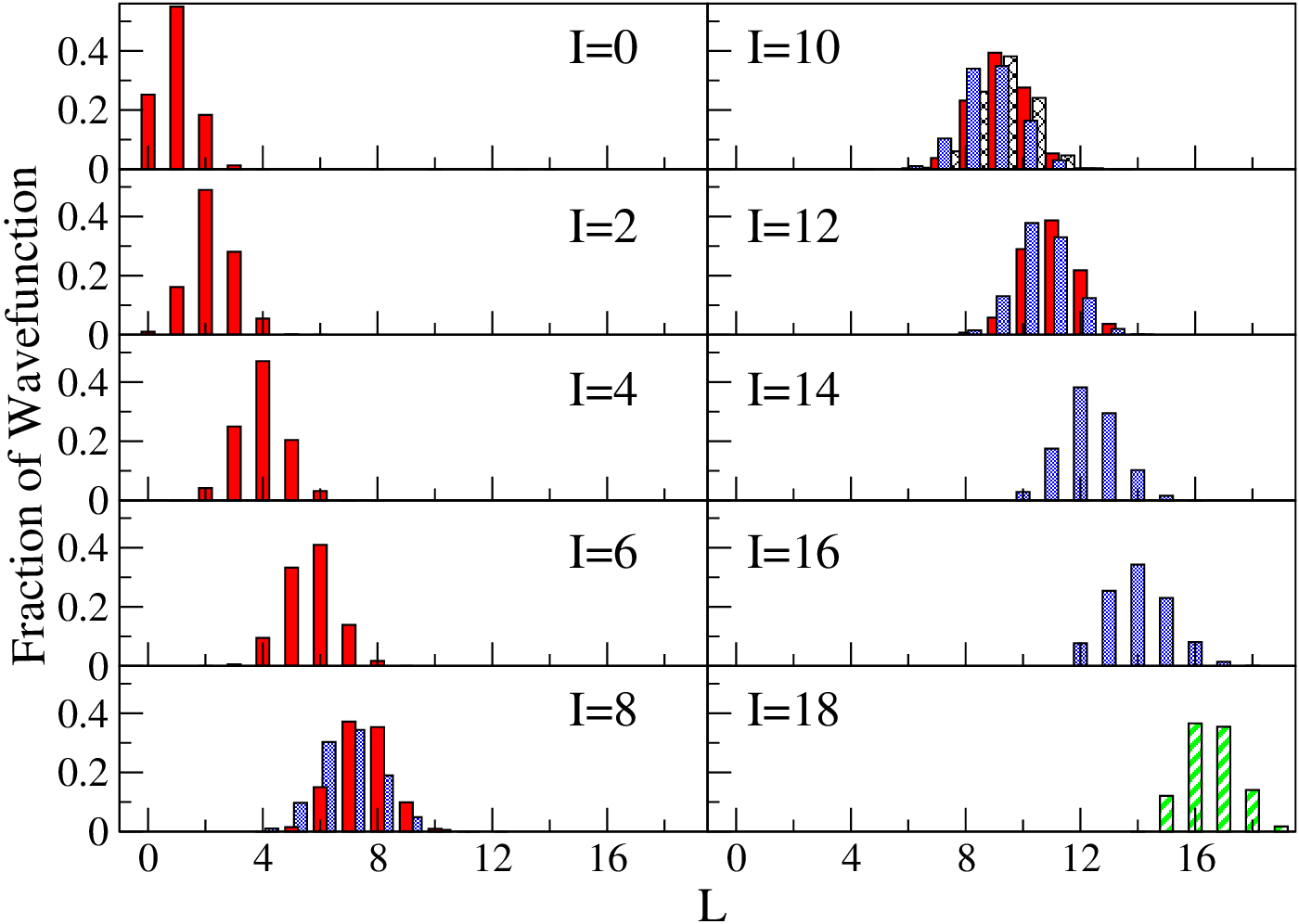}
\caption{(Color online) 
Decomposition of wavefunctions of $^{48}$Cr into components of total $L$ (orbital angular momentum).  The fill (and color) scheme are matched to the levels shown in Fig.~\ref{cr48spect}, 
i.e., (red) solid bars (lower sub-band), (blue) dotted (upper sub-band), and (black) cross-hatched, and (green) striped, intruder levels.  Here and throughout we superimpose levels which have the same $I$ but which belong to 
different sub-bands.}
\label{cr48Ldecomp}
\end{figure}

Modern computers allow us to carry out large scale calculations previously unimaginable. The $M$-scheme dimension for $^{48,49,50}$Cr in the 
$1p$-$0f$ valence space are 2 million, 6 million, and 14.6 million, respectively, but fully converged low-lying states can be computed in a matter of minutes on a 
laptop, and leadership-class configuration-interaction calculations have basis dimensions of the order of $10^{10}$.  This begs the question: do we really need that many 
numbers?

One attempt to simplify the description of nuclei is through \textit{dynamical symmetries}, where the Casimirs of a group commute with the 
nuclear Hamiltonian; then the eigenstates of the Hamiltonian will also be eigenstates of the Casimirs of the group, and one can just choose a basis within a 
single irreducible representation (irrep) of the group \cite{chen1989group,talmi1993simple,rowe2010fundamentals}, which is the smallest possible subspace where all group elements are block-diagonal. 
(The simplest, though still nontrivial, example of this would be a $J$-scheme basis, where the 
states have fixed total angular momentum $J$ rather than $M$. $J$-scheme bases are an order of magnitude smaller than $M$-scheme bases, but because 
each $J$-scheme state is a linear combination of $M$-scheme states, computing matrix elements is correspondingly more difficult and the Hamiltonian matrix 
is significantly denser.) 
The most prominent choice is the group SU(3), from which rotational bands arise naturally \cite{elliott1958collective,harvey1968nuclear}, or its extension 
the symplectic group Sp(3,R).  
 We loosely say we decompose the wavefunctions into group irreps, although in our SU(3) and SU(4) examples we use only one Casimir operator for the 
decomposition, and hence technically we in those cases we are combining results from different irreps.  In principle one could fully decompose into true irreps, but we chose not to, partly
to avoid in using three-body Casimirs for SU(3) as well as to keep our already busy figures become less readable.

Alas, it has long been known that the nuclear force, in particular the spin-orbit \cite{rochford1988survival,escher1998pairing,PhysRevC.63.014318} and pairing \cite{Bahri1995171} 
components, strongly mixes SU(3). But not all is lost: 
While the wavefunctions are distributed or \textit{fragmented} across many irreps, in many cases the patterns are strongly coherent and consistent across 
members of a band \cite{rochford1988survival,PhysRevC.63.014318}.  This 
is the concept of \textit{quasi-dynamical} symmetry \cite{PhysRevC.58.1539,rowe2000quasi,bahri20003} and helps to explain why SU(3) dynamical symmetry works well phenomenologically 
even though it fails microscopically. 

To illuminate quasi-dynamical symmetry, we decompose a wavefunction into subspaces labeled by Casimir eigenvalues.
Given a wave function $| \Psi_i \rangle$, which is an eigenstate of the nuclear many-body Hamiltonian (\ref{nucl_ham}), 
and a group Casimir $\hat{C}$ with eigenpairs
\begin{equation}
\hat{C}| z , \alpha \rangle = g(z) |  z, \alpha \rangle \label{casimir_eigen}
\end{equation}
where $z$ is a quantum number or numbers labeling subspaces of the group (for example, for SU(2) $I$ is a quantum number and $g(I) = I(I+1)$ ; note that, 
for consistency with many past papers on backbending, we use $I$ rather than $J$ for nuclear angular momentum) and 
$\alpha$ labels distinct states in the subspace, that is, solutions of (\ref{casimir_eigen}) degenerate in $g(z)$,  we want to find the fraction ${\cal F}(z)$ 
of the wave function $| \Psi_i \rangle $ in the subspace labeled by $z$, that is,
\begin{equation}
{\cal F}(z) = \sum_{\alpha \in z} \left | \langle z, \alpha | \Psi_i \rangle \right |^2.
\end{equation}
Luckily, there is an efficient method to find ${\cal F}(z)$ using the Lanczos algorithm \cite{PhysRevC.63.014318,PhysRevC.91.034313} that does not require finding all states in the irrep. 
This method only finds the magnitude in each subspace, not the phase.
 In the next section we plot ${\cal F}(z)$, 
 the fraction of the wavefunction in the subspace labeled by $z$, 
 versus either $z$ ( or $g(z)$, in the case of SU(3) and SU(4), where $z$ represents
 several labels) as bar graphs for states along the yrast band.

\begin{figure}
\centering
\includegraphics[scale=0.60,clip]{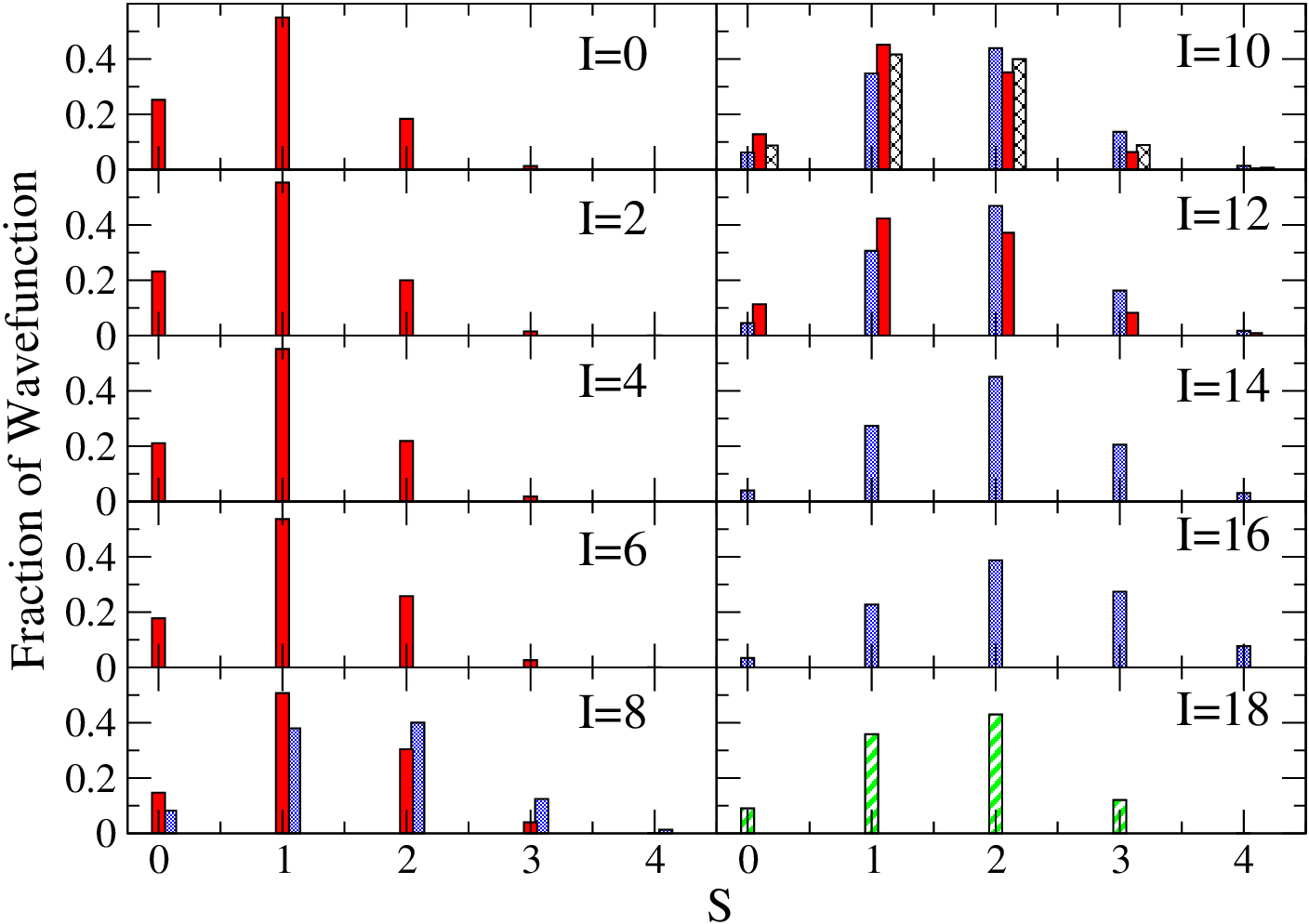}
\caption{(Color online) 
Decomposition of wavefunctions of $^{48}$Cr into components of total $S$ (spin).  The fill (and color) scheme are the same as in Fig.~\ref{cr48Ldecomp}.}
\label{cr48Sdecomp}
\end{figure}

The group Casimirs we use are: total orbital angular momentum $\hat{L}^2$ labeled by $L$; total spin $\hat{S}^2$ labeled by $S$; 
and the two-body Casimirs of SU(3) and SU(4).

The irreps of SU(3) are labeled by the quantum numbers  $\lambda$ and $\mu$ via their Young tableaux \cite{talmi1993simple}, and 
which can be interpreted in terms of the standard deformation parameters $\beta$ and $\gamma$ (see Figure 2 in Ref.~\cite{odz88} or Figure 1 in Ref.~\cite{Bahri1995171}).
We use only the two-body Casimir,
\begin{equation}
C_2(SU(3)) = \frac{1}{4} \left( \vec{Q} \cdot \vec{Q} + 3 L^2 \right),
\label{Csu3}
\end{equation}
where
\begin{equation}
Q_m= \sqrt{\frac{4\pi}{5}} \left ( \frac{r^2}{b^2} Y_{2m}(\Omega_r)+ b^2 p^2 Y_{2m}(\Omega_p) \right ),
\end{equation}
the (dimensionless) so-called Elliott quadrupole operator, whose matrix elements are nonzero only within a major harmonic oscillator shell; here 
$\Omega_r$ and $\Omega_p$ refer to the standard angles $\theta, \phi$ in spherical coordinates for the position and momentum vectors, respective. 
This Casimir has eigenvalues $\lambda^2 + \lambda \mu + \mu^2 + 3\lambda + 3\mu$ (in the above $b$ is the harmonic oscillator length parameter). 
One could distinguish between different combinations of $\lambda$ and $\mu$ by including the third-order Casimir, which is numerically more challenging. 
We discuss interpretation of the SU(3) decomposition in terms of deformation in Section \ref{deformed}.

Wigner suggested \cite{PhysRev.51.106,hecht1969wigner}  looking for an $SU(4)$ symmetry built upon 
$SU_S(2) \times SU_T(2)$, sometimes called a \textit{supermultiplet}.  
  The irreps of SU(4) are labeled by the quantum numbers $P,P^\prime$, and $P^{\prime \prime}$, which arise from the Young tableaux \cite{talmi1993simple,hecht1969wigner},
found by the Casimir operator
 \begin{equation}
 C_2(SU(4)) = \vec{S}^2 + \vec{T}^2 + 4 \sum_{i,j} (\vec{S}_i \cdot \vec{S}_j ) (\vec{T}_i \cdot \vec{T}_j ) 
 \end{equation}
 where the sum is over particles labeled by $i,j$, and 
 which has eigenvalues  \cite{talmi1993simple,hecht1969wigner},
 \begin{equation}
 P(P+4) +P^\prime(P^\prime+2) +\left( P^{\prime \prime}\right)^2
 \end{equation}
 In the highest weight states, $P=S$ and $P^\prime=T$. 
Despite its early history, SU(4) has recently been neglected, in part because it is badly broken in nuclei, for example in the $sd$ and $pf$ shells \cite{PhysRevC.47.623}. 
It has been primarily investigated in its role in the Wigner energy \cite{poves1998pairing}. Although we confirm  breaking of SU(4), we also demonstrate strong quasi-dynamical symmetry.

\begin{figure}
\centering
\includegraphics[scale=0.60,clip]{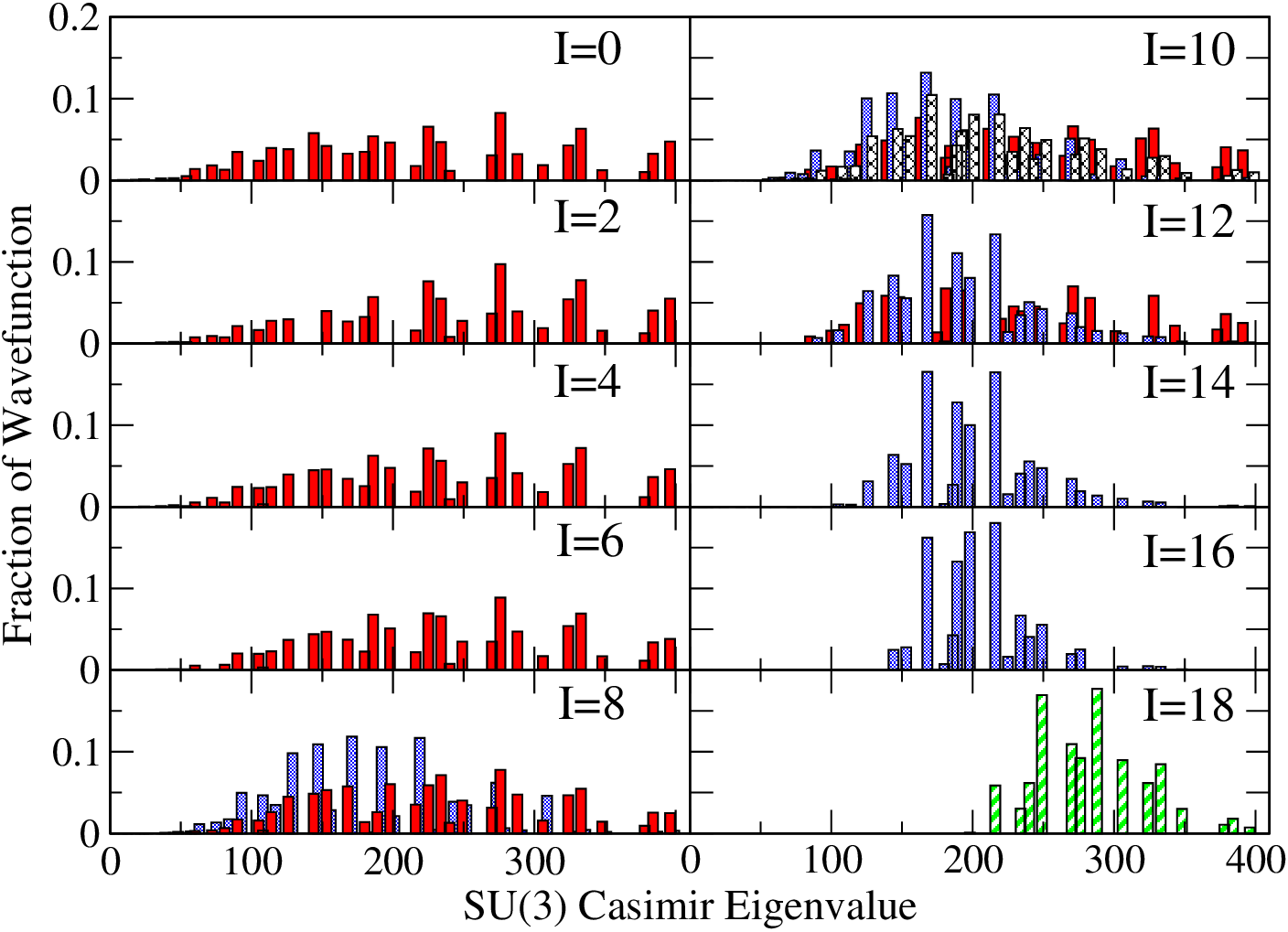}
\caption{(Color online) 
Decomposition of wavefunctions of $^{48}$Cr into SU(3) irreps, labeled by eigenvalues of the two-body SU(3) Casimir (see text for definition).  The fill (and color) scheme are the same as in Fig.~\ref{cr48Ldecomp}.}
\label{cr48su3}
\end{figure}

Group decompositions of the wavefunctions are of course not experimentally observable.  Prior work, however, in $L$- and $S$-decomposition comparing 
phenomenological and \textit{ab initio} calculations demonstrated remarkable consistency \cite{PhysRevC.91.034313}.  

\section{Results}

Throughout we attempt as much as possible to use a consistent labeling scheme of levels, e.g for  levels in the lower sub-band we use (red) solid circles for the 
excitation energies and (red) solid bars for the decomposition; for levels in the upper sub-band we use (blue) dotted triangles for excitation energies and (blue) dotted bars 
for decomposition; and finally for `intruder' states, that is, levels which do not belong to either the upper or lower sub-bands, we use black `x's and black cross-hatched 
bands and (green) striped circles/bars.  In all of this we group together levels via quasi-dynamical symmetry, that is, by inspecting the decomposition 
into irreps.  Using group decomposition and quasi-dynamical symmetry, we attempt to extend members of a band beyond the yrast in order to identify band crossings; 
we were able to do this for $^{48,50}$Cr but not $^{49}$Cr.

Although we attempt to give a reasonable summary of the existing literature, for purposes of comparison we emphasize those whose interpretations mostly clearly 
can be illuminated by our calculations, namely those which focus on shape deformations, and less so on $K$ quantum numbers (the $J_z$ value in the 
intrinsic frame) and quasi-particle excitations which, while of course relevant, are harder to connect to our group 
decompositions. 


\subsection{$^{48}$Cr}

We begin with backbending in $^{48}$Cr \cite{PhysRevC.49.1347,Cameron1996266}.  Fig.~\ref{cr48spect} shows the spectrum, spaced by $I(I+1)$ so that rotational bands are linear and easily picked out. 
In fact we see here and for our other two isotopes that the yrast bands are not ideal rotors but positioned between vibrational (linear in $I$)  and rotational (quadratic in $I$).

\begin{figure}

\centering
\includegraphics[scale=0.60,clip]{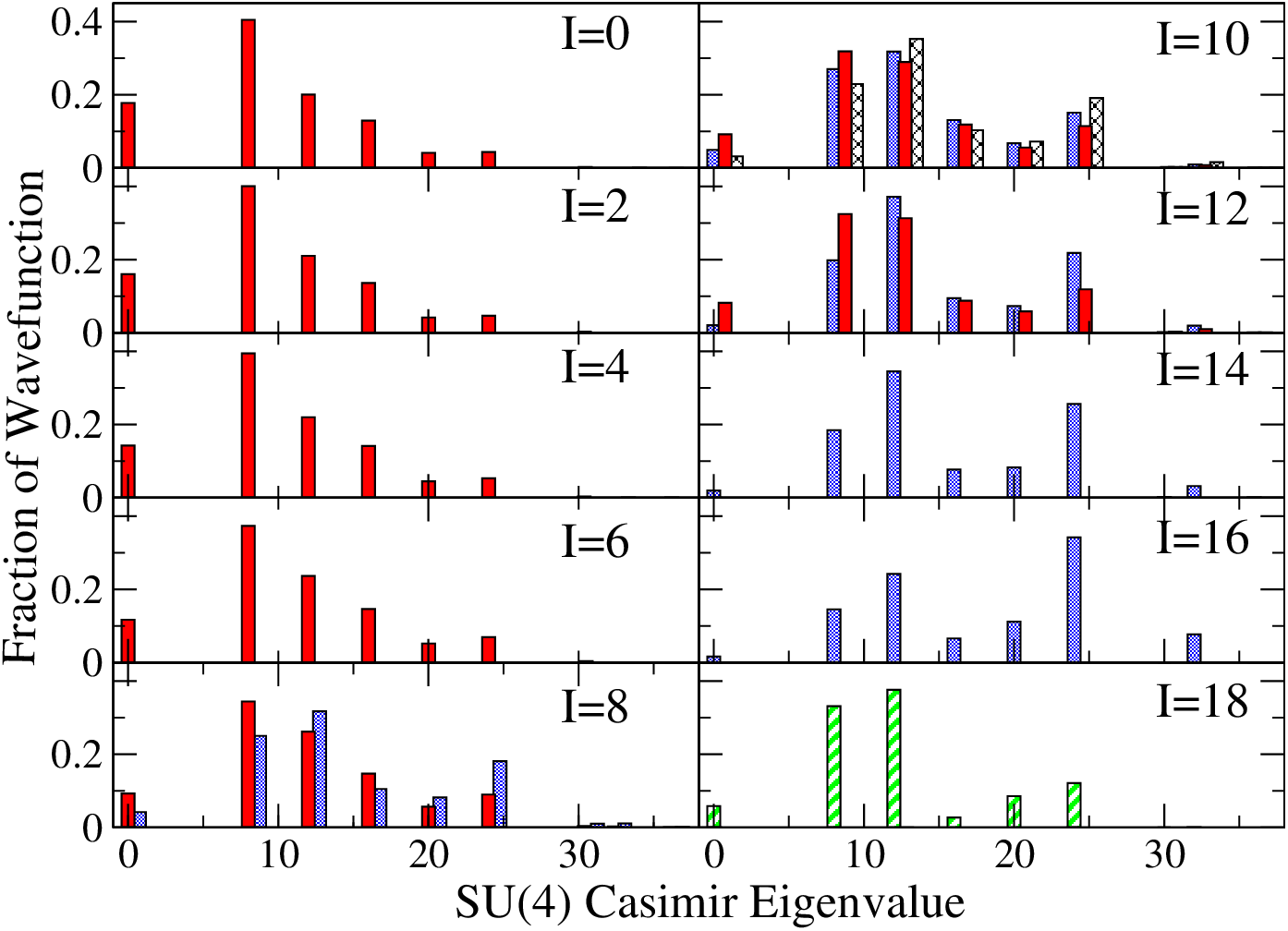}
\caption{(Color online) 
Decomposition of wavefunctions of $^{48}$Cr into SU(4) irreps, labeled by eigenvalues of the two-body SU(4) Casimir (see text for definition). The fill (and color) scheme are the same as in Fig.~\ref{cr48Ldecomp}.}
\label{cr48su4}
\end{figure}

Caurier \textit{et al.} \cite{PhysRevLett.75.2466} compared a cranked Hartree-Fock-Bogoliubov  (CHFB) calculation with the finite range Gogny force against a 
full $pf$-shell diagonalization. Both calculations yielded similar backbending and excellent agreement in $B$(E2) values, quadrupole and magnetic dipole moments, 
and orbital occupations;  the CHFB calculation showed an axially deformed rotor up to the backbend, while the yrast states after the backbend are more spherical and 
with the triaxiality parameter $\gamma$ less well-defined.   Because full space configuration-interaction (CI)  calculations do not have an intrinsic frame, the deformation cannot be computed directly, but Caurier \textit{et al.} argued that, given the good agreement between CI and CHFB in other quantities, the CHFB interpretation is likely robust. 

Later calculations support this picture. 
A subsequent CHFB calculation \cite{PhysRevC.58.2765} arrived at  similar results, i.e., consistent axial deformation up to the backbending, and then rapid transition to 
a spherical nucleus.  These authors emphasized the lack of a level crossing in the single-particle orbits, which is associated with backbending in heavier nuclides, 
and the importance of careful treatment of the residual interaction. 

Calculations with the ``projected shell model'' or PSM \cite{PhysRevLett.83.1922}, which uses a basis of deformed quasiparticle-quasihole states projected out with good 
angular momentum and particle number, also described the backbending 
of $^{48}$Cr in terms of a spherical band crossing a deformed band; furthermore, they identified \textit{two} crossings, the first around $I=6$, where a 2-quasiparticle (qp) band 
crosses the ground state 0-qp band, which does not show up as backbending, and the second, around $I=10$, where a 4-qp band crosses the 2-qp band. 


Finally the hybrid ``projected configuration interaction'' (PCI) \cite{PhysRevC.83.057303}, 
which is similar to the projected shell model but using deformed particle-hole states, that is, 
explicitly number-conserving, rather than quasiparticle-quasihole state, which are then projected out to good angular momentum and the Hamiltonian diagonalized  in this basis, found  results similar to that of Caurier 
\textit{et al.}. (Another germane difference is the PSM used a schematic interaction tuned to reproduce levels within their calculations, while the PCI uses semi-realistic 
shell-model interaction fitted within the full configuration space.)
  In particular they emphasized levels below the backbending are dominated by a single deformed intrinsic state, but not above the backbending.

Now we turn to our group decompositions for $^{48}$Cr.
The  $L$-decompositions, Fig.~\ref{cr48Ldecomp}, at first glance look like a intrinsic shape being spun up: the distribution of 
$L$ is similar for all the yrast states, though shifted up as total angular momentum $I$ increases.  But there are  subtleties. For example, the ground state is 
dominated by $L=1$, while the states $I=2,4,6,\ldots$ have their strength centering roughly around $L = I$.  Above the backbend at $I \approx 10$, this shifts;
now the strength centers roughly around $L \approx I -2$.  

This pattern is of course echoed in the $S$ decompositions (Fig.~\ref{cr48Sdecomp}): below the backbend, the 
decomposition is dominated by $S=1$, with some $S=0$ which decreases, and  $S=2$ which increases slightly, while after the backbend $S=2$ dominates with 
$S=1,3$ subdominant.  Of course, in this space the maximum $S$ is 4, which means when one reaches $I=18$ the minimum $L$ is $14$; this helps to explain 
the shifting pattern in the $L$ decomposition. 
Nonetheless, notice that the $I=18$ state is significantly different, particular in $S$. This is easily understood: the ground state band is predominantly $(0f_{7/2})^8$ 
\cite{PhysRevLett.75.2466} but the maximum angular momentum for that configuration is $I=16$.  

The SU(3)  decompositions, Figs.~\ref{cr48su3}, also show a pronounced change around the backbending. 
SU(3) is highly fragmented, as is well known for the $pf$ shell \cite{PhysRevC.63.014318}.
After the backbend, the distribution of SU(3) is much more narrow and in fact narrows further with increasing $I$. 
$K$-band termination may be contributing to this evolution, with some SU(3) $(\lambda, \mu)$ dropping out due to their maximum possible $L$ values.  On the other hand, 
the $L$- and $S$ decompositions do not change much within the uppper sub-band, until one reaches the termination of the $(0f_{7/2})^8$  configuration at $I=16$. 

Previous work  on SU(4)  only showed its fragmentation \cite{PhysRevC.47.623}, while we 
appear to be the first to demonstrate quasi-dynamical symmetry in SU(4) in the $pf$ shell, as in Fig.~\ref{cr48su4}. 
The SU(4) decomposition also changes 
dramatically at the backbend, although the spread does not evolve as it does so for SU(3).  
 Again the abrupt shifts at $I=18$  is easily interpreted as the termination of 
 the $(0f_{7/2})^8$  configuration band  at $I=16$.  
Interestingly, the change in the SU(4) decomposition at the backbend is most pronounced for $^{48}$Cr than for our other two nuclides. This is suggestive of 
studies investigating the relative role of isovector and isoscalar pairing in $N=Z$ and $N\neq Z$ nuclides, as in \cite{poves1998pairing}.

\begin{figure}
\centering
\includegraphics[scale=0.60,clip]{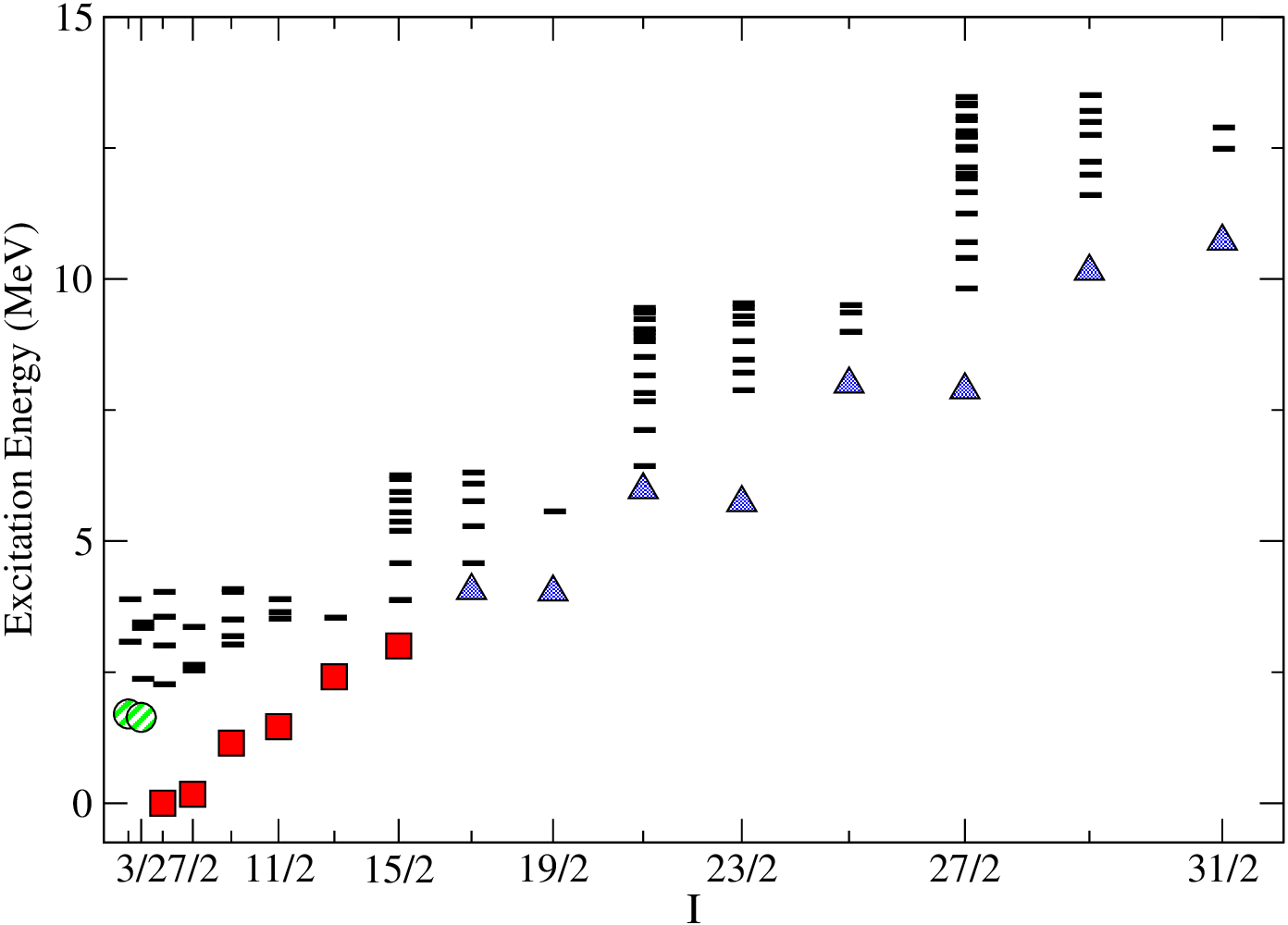}
\caption{(Color online) 
Calculated spectrum of $^{49}$Cr.
The $x$-axis (angular momentum $I$) is scaled as $I(I+1)$ so as to emphasize rotational bands.  The labeling of levels, i.e., 
red squares, blue triangles, and green circles, correspond to the same (initial) state as in Panel (b) in Fig.~\ref{BackBending}.  Bars indicate levels found in our 
calculation but which we do not decompose.
}
\label{cr49spect}
\end{figure}

By using the decompositions we were able to identify levels which are not part of the yrast band but which do appear to be continuations of the component sub-bands.
For example, we were able to trace the continuation of the lower sub-band up through $I=12$, as well as trace the upper sub-band down to $I=8$.  Futhermore we can see the actually yrast level at $I=10$, marked by 
an `x' in Fig.~\ref{cr48spect} and cross-hatched bars in Figs.~\ref{cr48Ldecomp}-\ref{cr48su4} belongs to 
neither the lower nor the upper sub-bands.

\subsection{$^{49}$Cr}

 Fig.~\ref{cr49spect} shows the spectrum of $^{49}$Cr spaced by $I(I+1)$.  
The yrast band of $^{49}$Cr has been measured up to $31/2^-$ \cite{PhysRevLett.79.4349,PhysRevC.58.808}, which is the highest angular momentum we calculate. 
It was previously calculated in the full $pf$ model space using shell-model CI \cite{PhysRevC.55.187}, where the authors explicated the results in terms of Nilsson diagrams and detailed effects 
of the residual interaction; other calculations emphasize the role of $K$-bands and quasi-particle excitations of the intrinsic state 
\cite{velazquez2001band,PhysRevC.71.054316,PhysRevC.73.044327}.

\begin{figure}
\centering
\includegraphics[scale=0.60,clip]{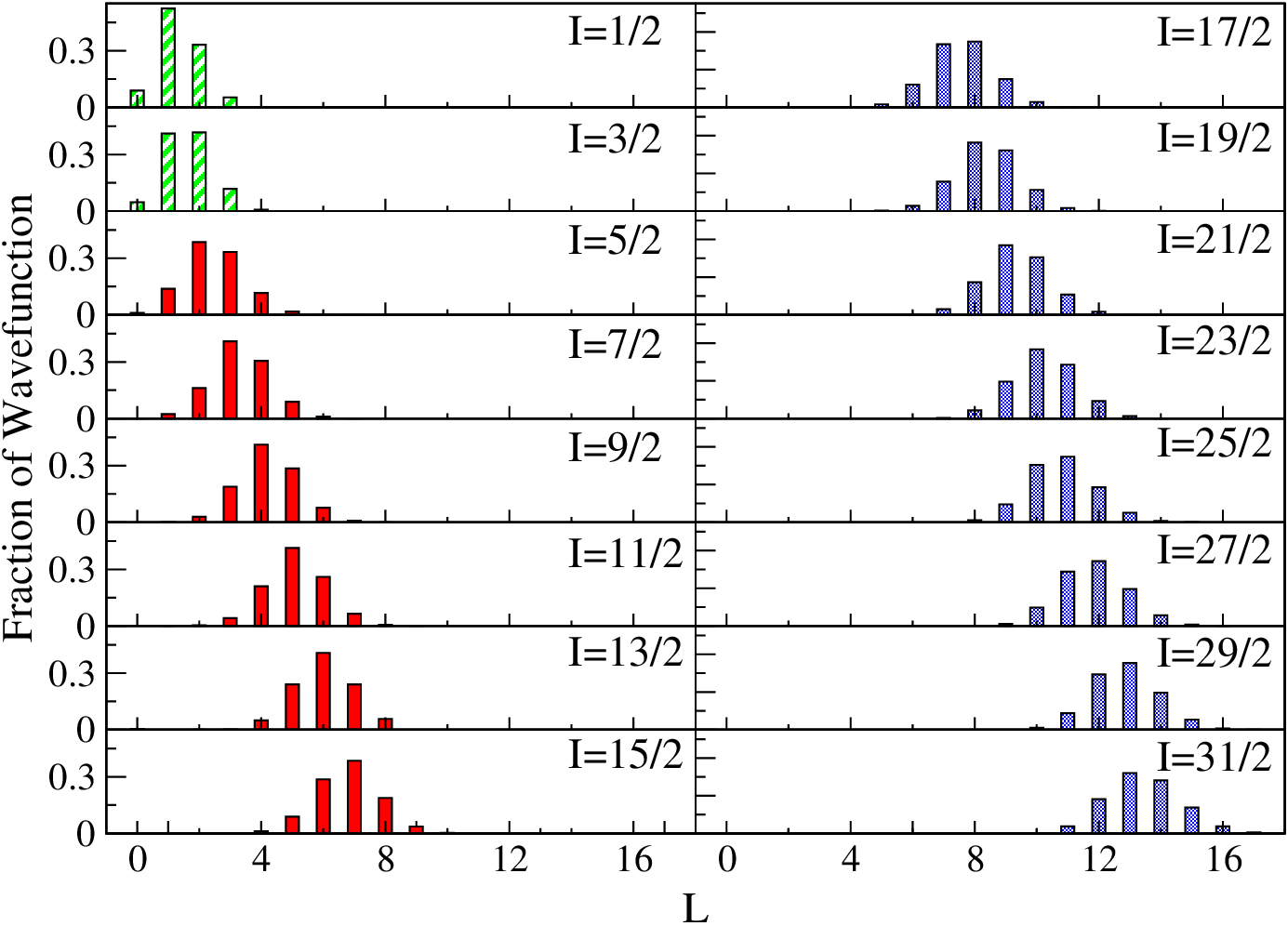}
\caption{(Color online) 
Decomposition of wavefunctions of $^{49}$Cr into components of total $L$ (orbital angular momentum). Much like Fig.~\ref{cr48Ldecomp}, 
 the fill (and color) scheme are matched to the levels shown in Fig.~\ref{cr49spect}, 
i.e., (red) solid bars (lower sub-band), (blue) dotted (upper sub-band), and (green) striped, the lowest $I=1/2,3/2$ which technically are not part of the yrast line.  
}
\label{cr49Ldecomp}
\end{figure}

As with all three of our 
nuclides, the $L$ decompositions, Fig.~\ref{cr49Ldecomp}, increase steadily with $I$;  similar to what we saw with $^{48}$Cr, below the the $L$-decompositions for 
each angular momentum $I$ centers around $L \approx I-1/2$, while in the upper sub-band it centers around $L\approx I-3/2$.

\begin{figure}
\centering
\includegraphics[scale=0.60,clip]{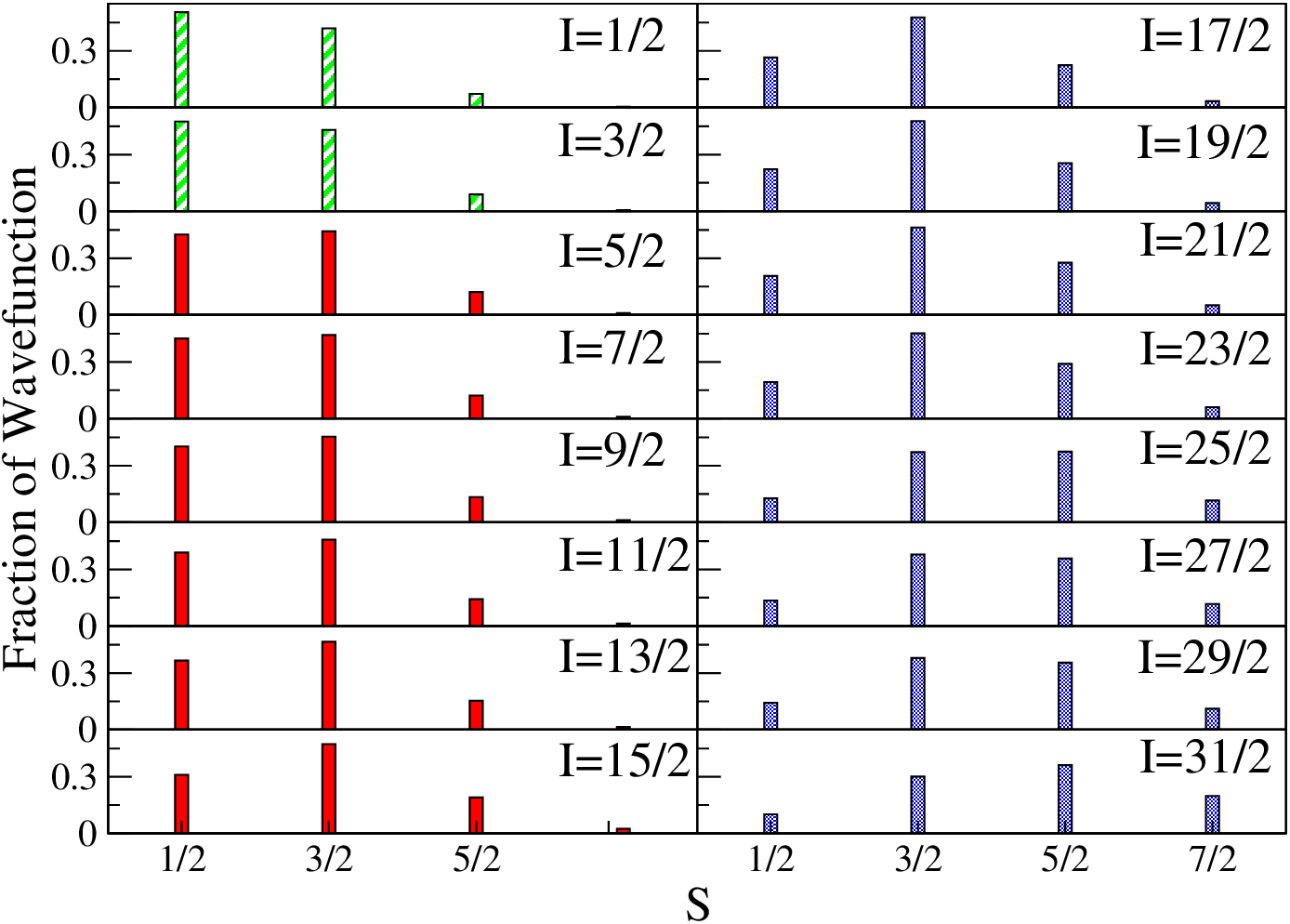}
\caption{(Color online) 
Decomposition of wavefunctions of $^{49}$Cr into components of total $S$ (spin).  The fill (and color) scheme are the same as in Fig.~\ref{cr49Ldecomp}. }
\label{cr49Sdecomp}
\end{figure}

The spin decompositions, Fig.~\ref{cr49Sdecomp} show strong (but distinct) quasi-dynamical symmetry below and above the backbend, and could be approximated by taking the spin 
decompositions of $^{48}$Cr and shifting up by 1/2 unit of angular momentum ( the $L$-decomposition also strongly parallel that of $^{48}$Cr): below the backbend  the yrast band is dominated by $S=1/2, 3/2$, while above the 
backbend $S=3/2, 5/2$ dominate.

\begin{figure}
\centering
\includegraphics[scale=0.60,clip]{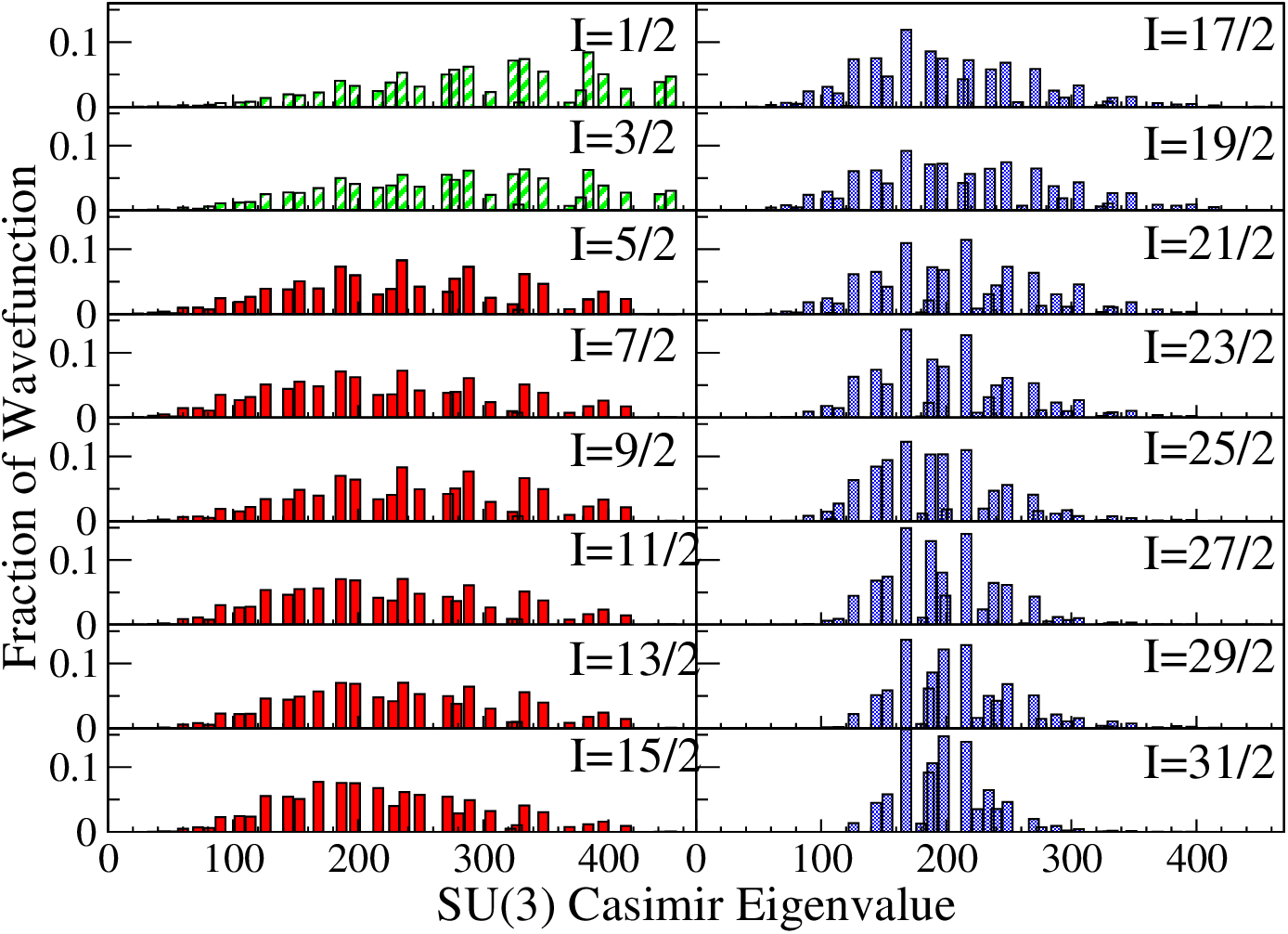}
\caption{(Color online) 
Decomposition of wavefunctions of $^{49}$Cr into SU(3) irreps.  See  text for the definition of the SU(3) Casimir.  The fill (and color) scheme are the same as in Fig.~\ref{cr49Ldecomp}.}
\label{cr49su3}
\end{figure}

Also like  $^{48}$Cr, the SU(3) decomposition of $^{49}$Cr, Fig.~\ref{cr49su3}, is relatively coherent below the backbend,  while above the backbend the distribution becomes 
narrower and has more pronounced evolution. 

\begin{figure}
\centering
\includegraphics[scale=0.60,clip]{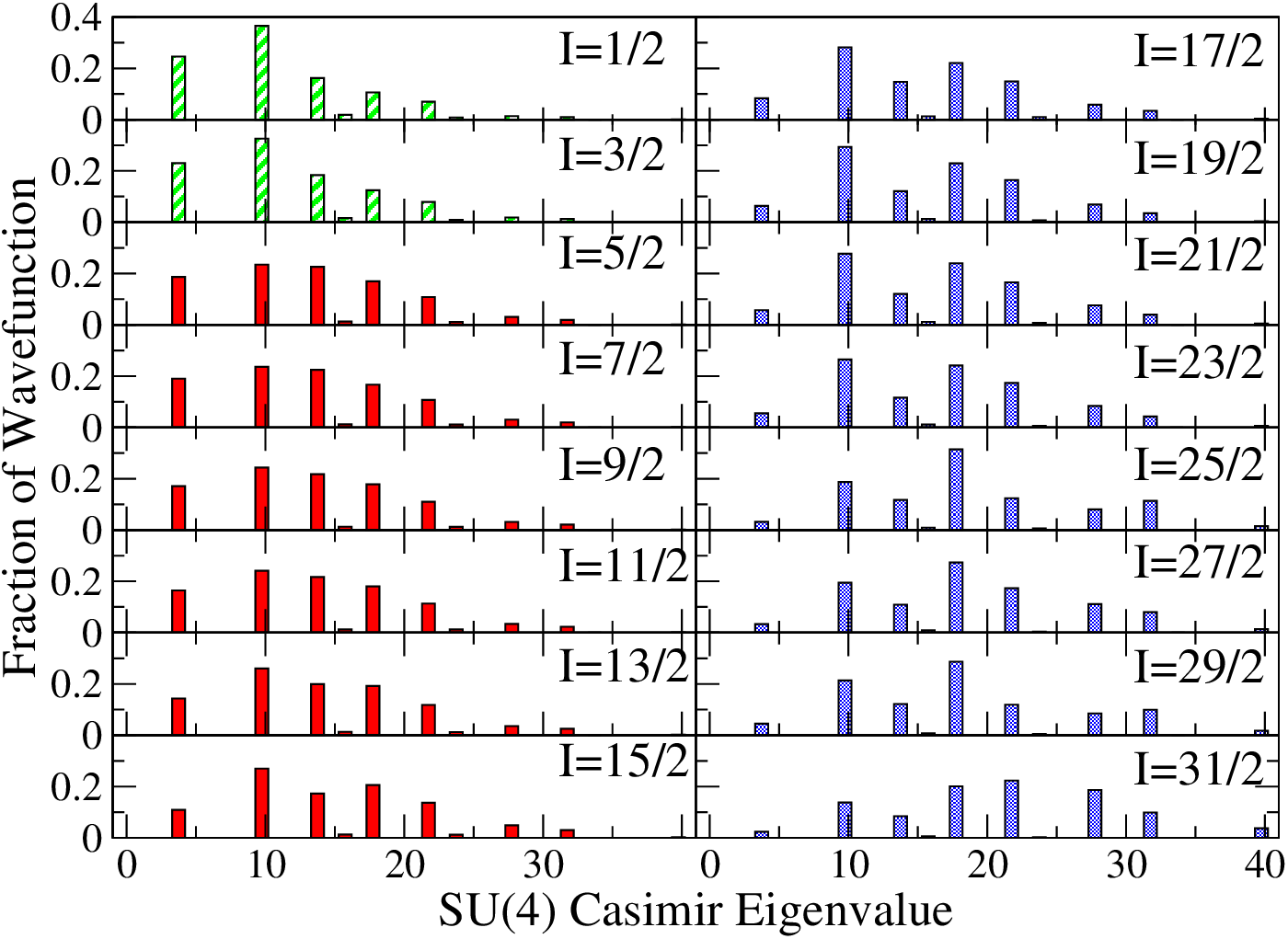}
\caption{(Color online) 
Decomposition of wavefunctions of$^{49}$Cr into SU(4) irreps.  See  text for the definition of the SU(4) Casimir.  The fill (and color) scheme are the same as in Fig.~\ref{cr49Ldecomp}.}
\label{cr49su4}
\end{figure}

Fig.~\ref{cr49su4}  shows strong quasi-dynamical symmetry in SU(4), especially in the lower sub-band, but with 
significant coherence in the upper band as well; while there is a definite change across the backbend, it is not as dramatic as for $^{48}$Cr.
Here we were not able to identify continuations of the sub-bands beyond their locations on the yrast band.

In our figures we include the low-lying $I=1/2, 3/2$ levels which, though part of the yrast band, are not the yrast band heads; 
in the $S$ and SU(4) decompositions they clearly are grouped with the rest of the low-lying yrast levels, but they have nontrivial 
differences in the other decompositions, most markedly in SU(3).

\subsection{$^{50}$Cr}

\begin{figure}
\centering
\includegraphics[scale=0.60,clip]{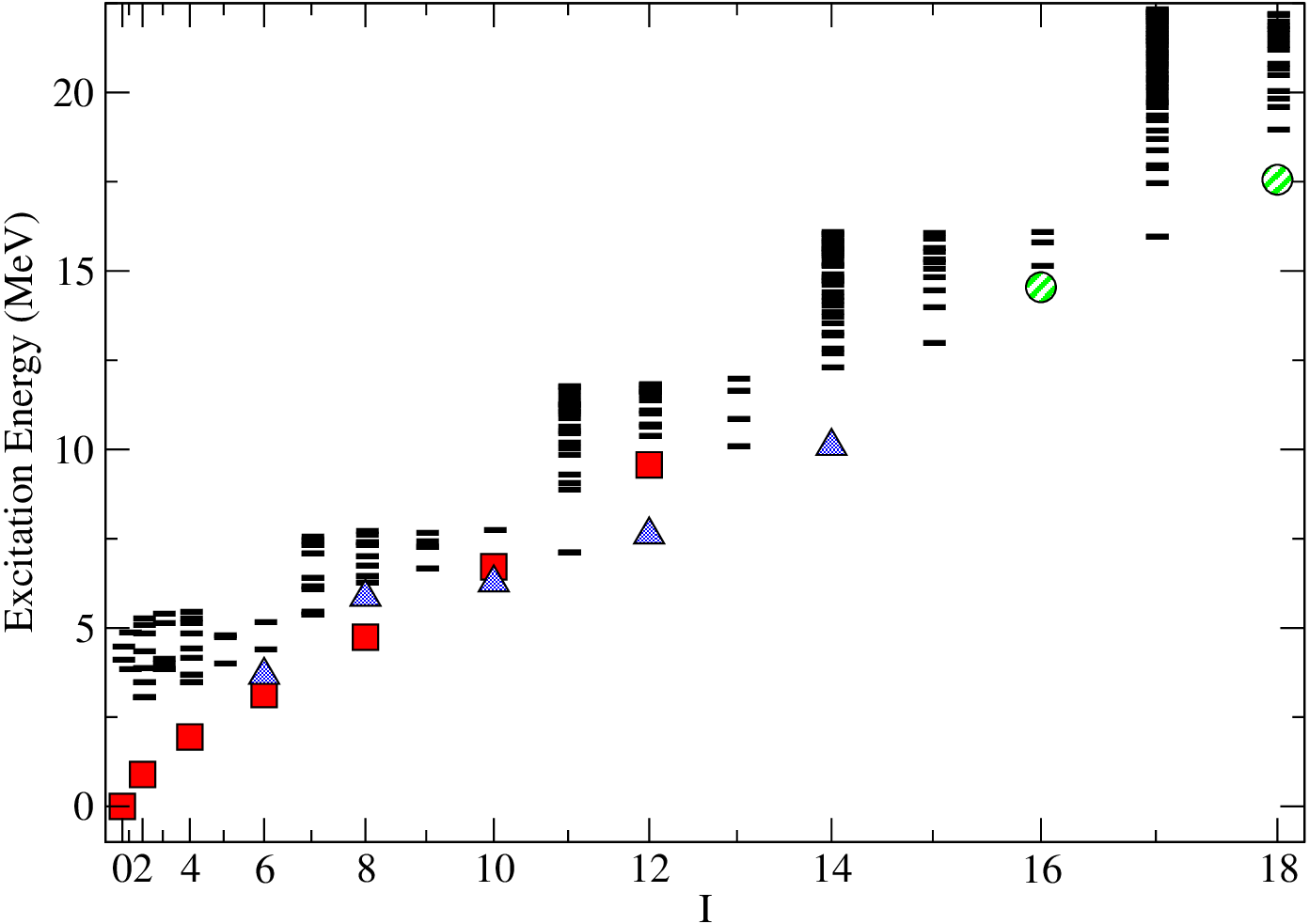}
\caption{(Color online) 
Calculated spectrum of $^{50}$Cr.  The $x$-axis (angular momentum $I$) is scaled as $I(I+1)$ so as to emphasize rotational bands.  The labeling of levels, i.e., 
(red) squares, (blue) triangles, and (green) circles, correspond to the same (initial) state as in Panel (a) of Fig.~\ref{BackBending}.  
 Bars indicate levels found in our 
calculation but which we do not decompose.}
\label{cr50spect}
\end{figure}

The yrast band of $^{50}$Cr has been measured up to $I^\pi =18^+$ \cite{PhysRevC.56.1313,PhysRevC.58.808,PhysRevC.66.021302}, as shown in Fig.~\ref{cr50spect}, with backbending seen around $I \approx 10$ and a second
backbending around $I\approx 16$ which is easily interpreted as the terminus of levels generated within the $(0f_{7/2})^{10}$ configuration. 
The origin of the change at the backbending is somewhat unclear within CI calculations; Mart\'inez-Pinedo \textit{et al} \cite{PhysRevC.54.R2150} interpret it as a shift from strongly prolate to 
weakly oblate, similar to what is seen in $^{48}$Cr, yet Zamick \textit{et al}, looking at the sign of the quadrupole moments in just the $(0f_{7/2})^{10}$ configuration space 
\cite{PhysRevC.53.188}, argue instead the upper sub-band could belong to a high-$K$ prolate band. 

Similar to the work on $^{48}$Cr \cite{PhysRevLett.75.2466},  calculations using the configuration-interaction (CI) shell model were compared directly with 
cranked Hartree-Fock-Bogoliubov calculations \cite{PhysRevC.54.R2150}, and with similar  results: both CI and CHFB showed backbending at $I\approx 10$ and
$I \approx 16$; the latter is where pure $(0f_{7/2})^{10}$ configurations must terminate.  In particular they find $^{50}$Cr to be axially symmetric and prolate below $I\approx 10$, 
afterwhich it becomes oblate and weakly triaxial, until it reaches $I\approx 16$ where, again at the termination of the $(0f_{7/2})^{10}$ configuration it becomes strongly triaxial. 

\begin{figure}
\centering
\includegraphics[scale=0.60,clip]{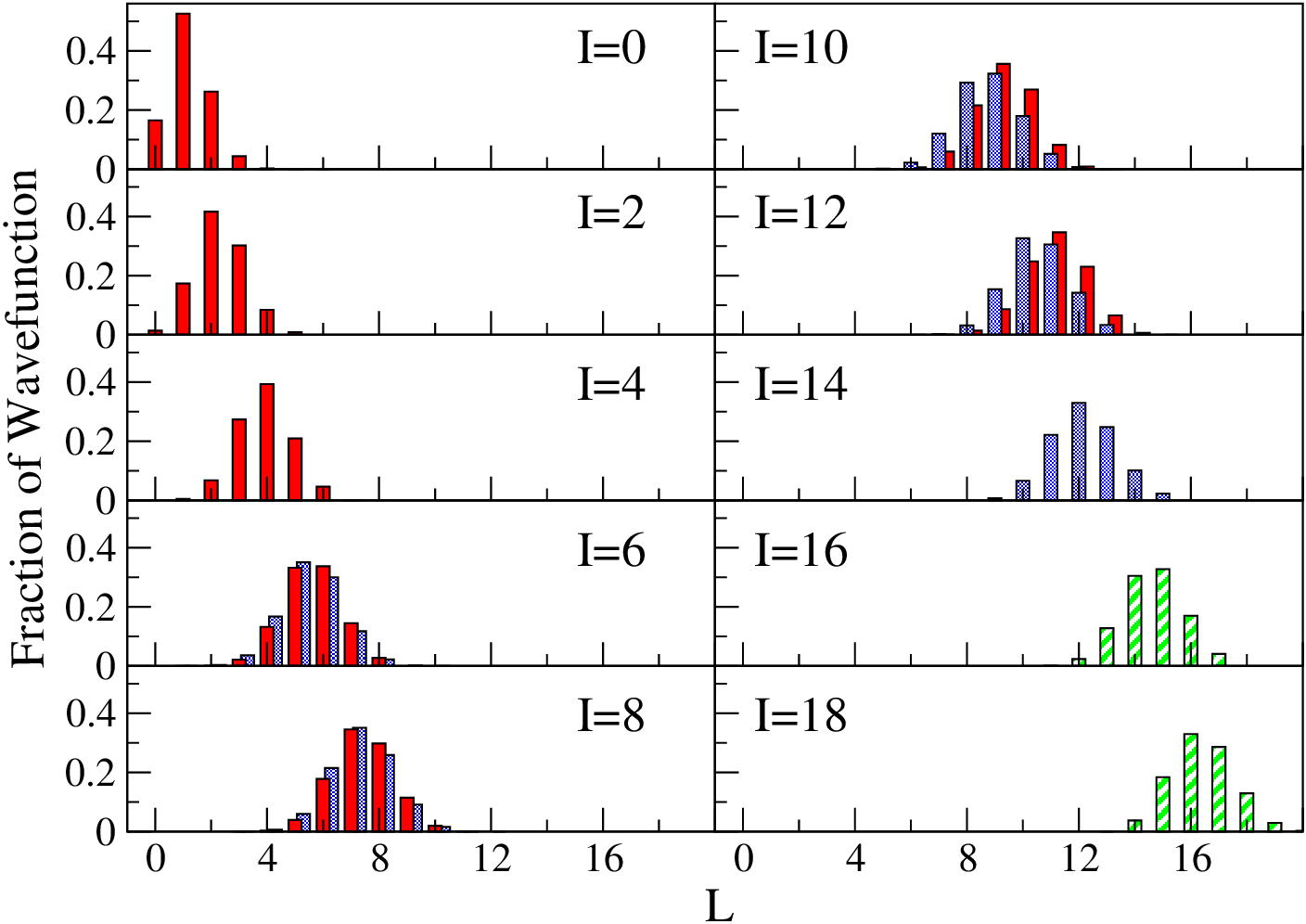}
\caption{(Color online) 
Decomposition of wavefunctions of  $^{50}$Cr into components of total $L$ (orbital angular momentum).  Decomposition of wavefunctions of $^{50}$Cr into components of total $L$ (orbital angular momentum). Much like Fig.~\ref{cr48Ldecomp}, 
 the fill (and color) scheme are matched to the levels shown in Fig.~\ref{cr50spect}, 
i.e., (red) solid bars (lower sub-band), (blue) dotted (upper sub-band), and (green) striped (`intruder,' that is, outside of the $(0f_{7/2})^{10}$ configuration space).  Here and throughout we superimpose levels which have the same $I$ but which belong to 
different sub-bands.}
\label{cr50Ldecomp}
\end{figure}

\begin{figure}
\centering
\includegraphics[scale=0.60,clip]{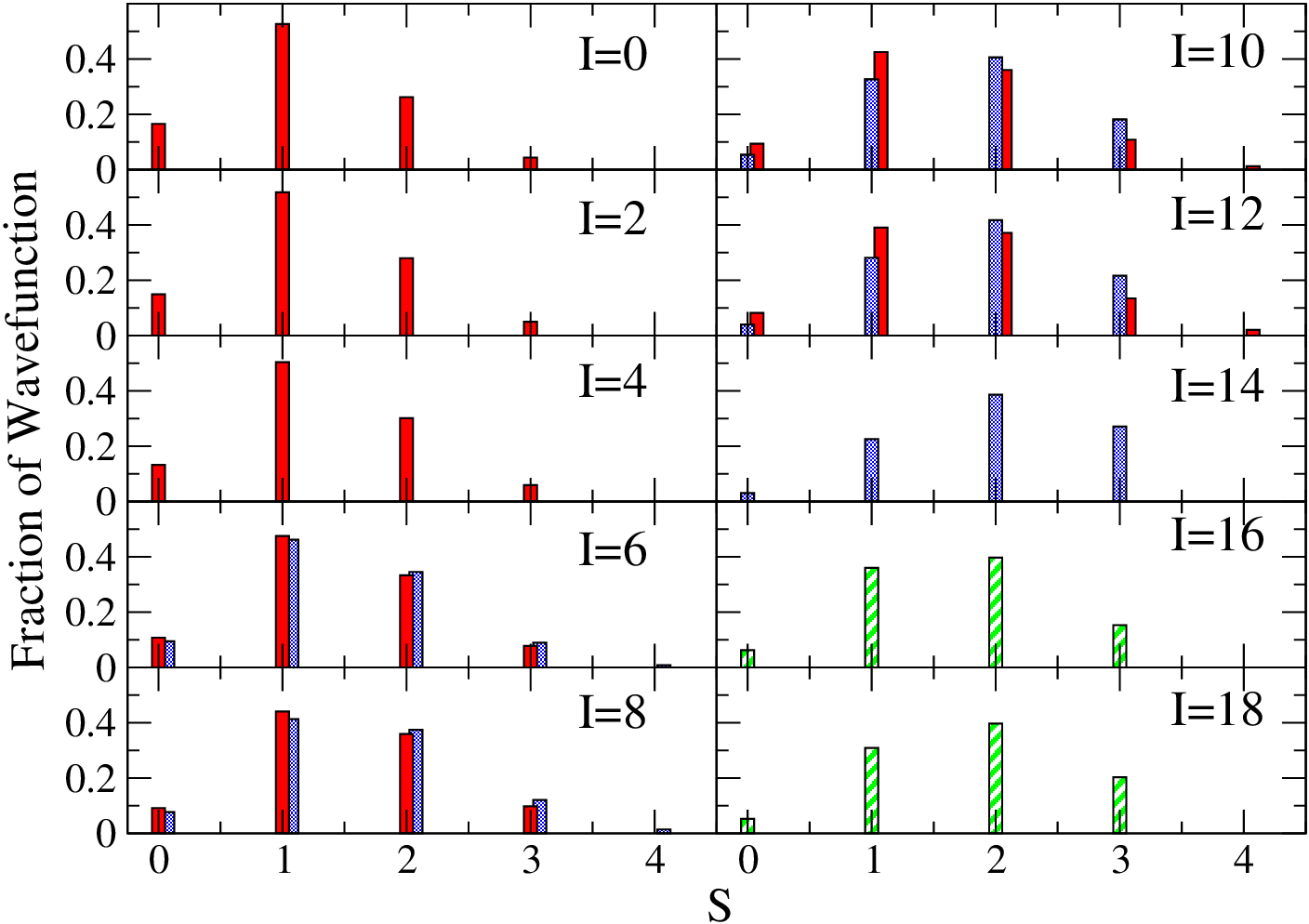}
\caption{(Color online) 
Decomposition of wavefunctions of  $^{50}$Cr into components of total $S$ (spin).  Fill (and color) scheme same as that of Fig.~\ref{cr50Ldecomp}.  }
\label{cr50Sdecomp}
\end{figure}

\begin{figure}
\centering
\includegraphics[scale=0.60,clip]{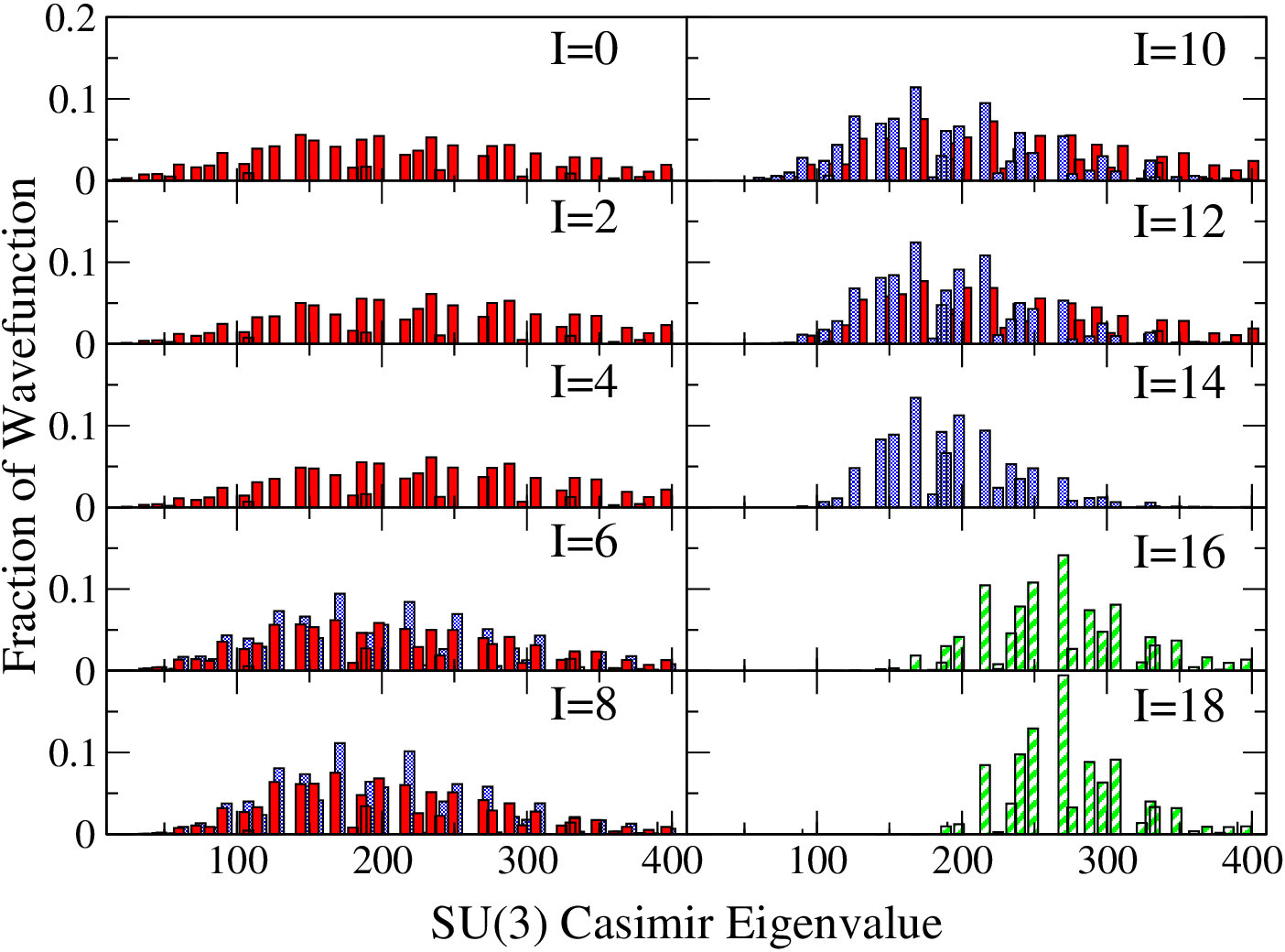}
\caption{(Color online) 
Decomposition of wavefunctions of $^{50}$Cr into SU(3) irreps.   Fill (and color) scheme same as that of Fig.~\ref{cr50Ldecomp}.}
\label{cr50su3}
\end{figure}

\begin{figure}
\centering
\includegraphics[scale=0.60,clip]{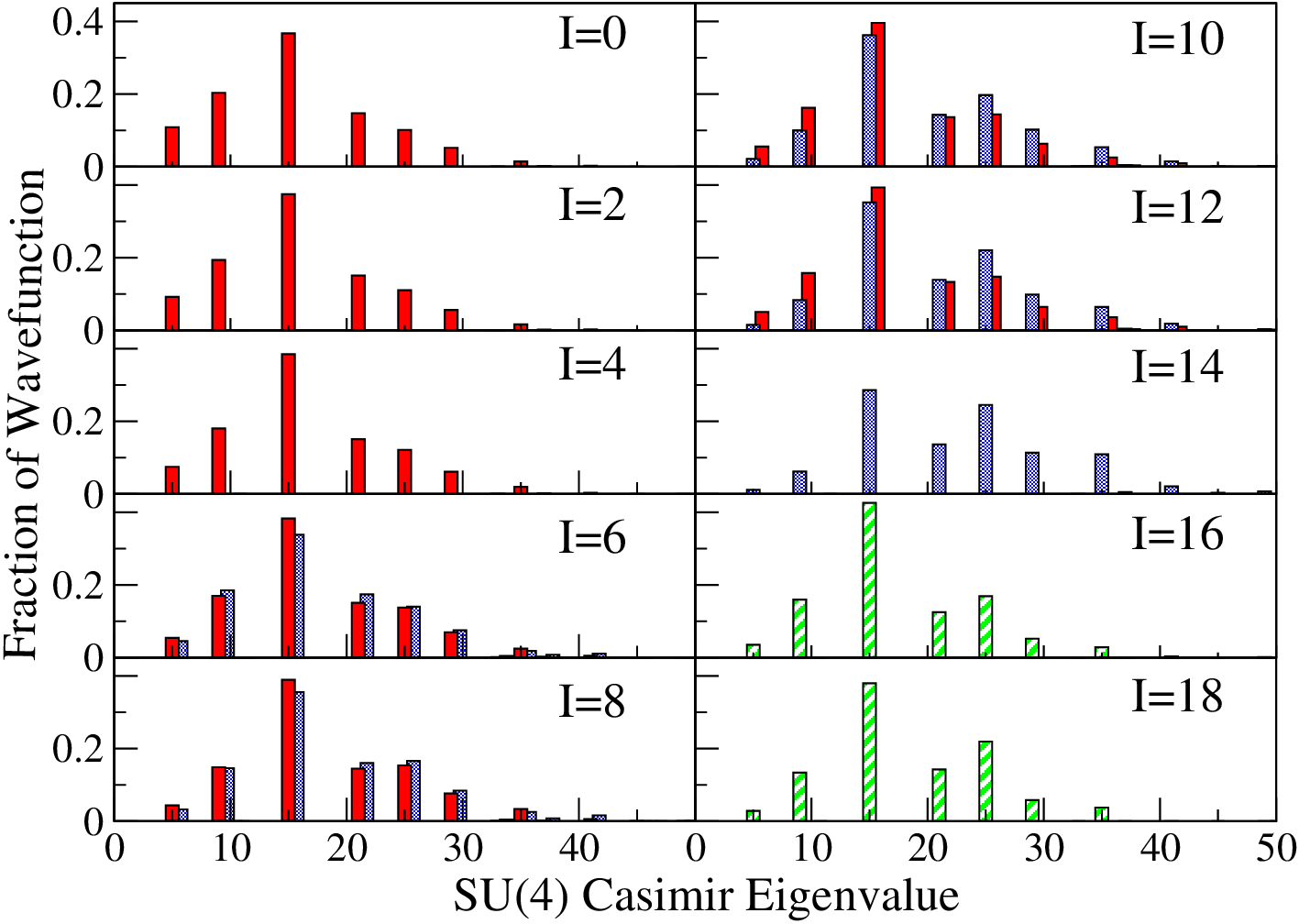}
\caption{(Color online) 
Decomposition of wavefunctions of $^{50}$Cr into SU(4) irreps.   Fill (and color) scheme same as that of Fig.~\ref{cr50Ldecomp}.}
\label{cr50su4}
\end{figure}

While the decomposition in $L$, shown in Fig.~\ref{cr50Ldecomp}, shows significant shifts at the two backbending points, the decompositions in $S$, Fig.~\ref{cr50Sdecomp}, and 
SU(4), Fig.~\ref{cr50su4}, are more subtle than for our other two nuclides: in the run-up to the backbend, at $I=6,8$, the decompositions of both sub-bands are nearly identical, but as $I$ increases 
up to and past the backbend at $I=12$, the decompositions of the upper sub-band shows a stronger evolution.
 Like the other nuclides, in the SU(3) decomposition, Fig.~\ref{cr50su3}, we see strong quasi-dynamical symmetry in the lower 
sub-band, with strong changes at the two backbends, and the fragmentation becoming more narrow.

\subsection{SU(3) and  deformations}
\label{deformed}

For a given state wholly in an SU(3) irrep labeled by $(\lambda,\mu)$ one can map it to a deformed shape and determine its deformation parameters 
$\beta$ and $\gamma$ ; in particular, the value of the two-body SU(3) Casimir is proportional to $\beta^2$  \cite{odz88}.
This has been used in prior work to examine SU(3) breaking by the pairing and spin-orbit forces \cite{Bahri1995171,escher1998pairing}. The broad fragmentations we see 
in SU(3) is similar to the broad distributions of $\beta$ and $\gamma$ values  in the presence of strong spin-orbit splitting in Figs.~2 and 3 of \cite{escher1998pairing}.

 It is therefore tempting to interpret our SU(3) decompositions as telling us something about deformation. 
By eye one can see, and we confirmed in detail,  the  expectation value of $C_2(SU(3))$ does not change much along the 
yrast line for each of our nuclides; by the above mapping this would suggest the average value of $\beta^2$ also remains near constant.  
 This, however, contradicts prior work using mean-field frameworks suggesting $^{48,49,50}$Cr are all
 strongly prolate, axially symmetric rotors below the backbend, while above the backbend they becomes nearly spherical and are less well-interpreted in terms of a single intrinsic shape 
\cite{PhysRevLett.75.2466,PhysRevC.58.2765,PhysRevC.73.044327,PhysRevC.83.057303,PhysRevC.73.044327,PhysRevC.54.R2150}.  
(Although we do not show it, we confirmed this behavior with a separate Hartree-Fock code using shell-model interactions.)

 It is important to note   that a deformed Slater determinant does not necessarily correspond to a single SU(3) irrep. Rather, it can be fragmented across many group irreps, as  previously 
 demonstrated in \cite{PhysRevC.66.034312}, where a projected Hartree-Fock state had a much stronger overlap with the full configuration-interaction ground state wavefunction than the highest-weight SU(3) state,
 driven   predominantly by the single-particle spin-orbit force.  
 
 We can provide a class of simple examples which show the mapping of SU(3) labels $(\lambda,\mu)$ to deformation can conflict with a simple mean-field picture.
 Consider a state which consists of a filled single-$j$ shell, for example, $^{48}$Ca where one fills the $0f_{7/2}$ shell with neutrons.  This is a single Slater determinant
 and is a manifestly spherical shape: the expectation value of the quadrupole tensor vanishes. 
 Yet if one decomposes it using the SU(3) two-body Casimir, it has only a $1\%$ fraction in the spherical $(\lambda,\mu)=(0,0)$ irrep; the rest of the 
 wavefunction is broadly spread across many SU(3) irreps. This result is not unique to $^{48}$Ca, but occurs whenever one fills a $j$-shell but not its spin-orbit partner. 
 The fact that one has large SU(3) mixing is not  surprising, given  the  spin-orbit splitting, but it also  suggests a picture of deformation can depend strongly upon whether 
 determined from a mean-field solution or from an SU(3) decomposition.

\section{Conclusions and acknowledgements}

In order to illuminate  backbending in  chromium isotopes, we carried out group  decomposition of shell model CI wavefunctions, using total orbital angular momentum $L$,
total spin $S$,  and the two-body Casimir operators of SU(3) and SU(4).   
We saw strong quasi-dynamical symmetry in all cases, often with a significant shift in the fragmentation as one crosses from the lower to the upper sub-band. 
Above the backbend the SU(3) distribution show the largest evolution with increasing $I$,
 a narrowing of the distribution but with a nearly constant average. 
On one hand large expectation values of the  SU(3) two-body Casimir eigenvalues suggest persistent large deformation, but 
 mean-field calculations consistently depict the yrast states at high $I$ have decreasing deformation. 
We note this clash of deformation pictures, that is, mean-field versus SU(3), can be found even in the very simple example of a simple spherical Slater determinant, a filled $j$-shell, which 
also has a broad distribution across many deformed SU(3) irreps. 

 In contrast, spin $S$ and SU(4) show less evolution in the sub-bands, both below and
 above the backbending. 
SU(4) shows the most pronounced shift in decomposition at the backbend in $^{48}$Cr, much less so in our other two nuclides;
nonetheless, we have demonstrated pervasive SU(4) quasi-dynamical symmetry in the $pf$ shell. 
Overall the $L$ decomposition simply shows a steady and coherent increase in angular momentum. 

Of course, the $pf$ shell space is limited and the GXPF1 interaction is phenomenological and heavily renormalized relative to the `real'  nuclear force. 
While there has been work decomposing \textit{ab initio} 
wavefunction for very light nuclei into SU(3) irreps, \cite{PhysRevLett.111.252501}, quasi-dynamical symmetry has not been deeply investigated in such calculations. 
We only note that one previous investigation, in the $L$ and $S$ decomposition only \cite{PhysRevC.91.034313} in $p$-shell nuclei, showed remarkable congruence between results from phenomenological 
and \textit{ab initio} interactions. 

While it would  be interesting to apply these same analyses to heavier nuclei with backbending, the fact that tractable model spaces for such nuclei general exclude 
spin-orbit partners makes exact decomposition impossible.  One could consider pseudospin, pseudo-SU(3), and other approximate symmetries, but this we also leave to future 
work.

This material is based upon work supported by the U.S. Department of Energy, Office of Science, Office of Nuclear Physics, 
under Award Number  DE-FG02-03ER41272.  We thank J. Escher, K. Launey, and P. van Isacker for stimulating discussions regarding the interpretation of deformation via SU(3) irreps.

\bibliographystyle{apsrev4-1}
\bibliography{johnsonmaster}
\end{document}